\documentclass[11pt]{article}
\usepackage{graphicx}

\newcommand{\BABARPubYear}    {06}

\newcommand{\BABARProcNumber} {062}
\newcommand{\SLACPubNumber} {11958}

\input pubboard/babarsym

\def\babar{\mbox{\slshape B\kern-0.1em{\smaller A}\kern-0.1em
    B\kern-0.1em{\smaller A\kern-0.2em R}}\xspace}

\newcommand{\Bxclnu}{\ensuremath{\Bb \rightarrow X_c \ell \bar{\nu}}\xspace}
\newcommand{\Bxulnu}{\ensuremath{\Bb \rightarrow X_u \ell \bar{\nu}}\xspace}

\def\Btodstlnu  {\ensuremath{\Bb \to \Dstar \ell \bar\nu}\xspace}

\newcommand{\Q}{\ensuremath{q^2}\xspace}

\long\def\inst#1{\par\nobreak\kern 4pt\nobreak
    {\it #1}\par\vskip 10pt plus 3pt minus 3pt}

\begin{document}
{\pagestyle{empty}

\begin{flushright}
SLAC-PUB-\SLACPubNumber \\
\babar-PROC-\BABARPubYear/\BABARProcNumber \\
July 18, 2006 \\
\end{flushright}

\par\vskip 4cm

\begin{center}
\Large \bf Measurement of the CKM Matrix Elements \Vcb and \Vub at the B-factories
\end{center}
\bigskip

\begin{center}
\large 
W. Menges\\
Department of Physics, Queen Mary, University of London \\
Mile End Road, London, E1 4NS, UK \\
(from the \lbabar\ Collaboration)
\end{center}
\bigskip \bigskip

\begin{center}
\large \bf Abstract
\end{center}
Recent results on inclusive and exclusive semileptonic \B decays from
\B-factories are presented. The impact of these measurements on the
determination of the CKM matrix elements \Vub and \Vcb is discussed.

\vfill
\begin{center}
Contributed to the Proceedings of the\\
14$^{th}$ International 
Workshop On Deep Inelastic Scattering (DIS 2006), \\
20-24 Apr 2006, Tsukuba, Japan
\end{center}

\vspace{1.0cm}
\begin{center}
{\em Stanford Linear Accelerator Center, Stanford University, 
Stanford, CA 94309} \\ \vspace{0.1cm}\hrule\vspace{0.1cm}
Work supported in part by Department of Energy contract DE-AC03-76SF00515.
\end{center}

\section{Introduction}

Semileptonic \B decays provide direct access to the CKM matrix
elements \Vub\ and \Vcb, whose ratio measures the side
of the Unitarity Triangle opposite the angle $\beta$.
Semileptonic \B decays also probe the structure of \B mesons.
Inclusive decays are sensitive to quantities such as the mass and
momentum distribution of the $b$ quark inside the \B meson, whereas
exclusive decays depend on form factors for specific final states. The
theory is quite advanced and the measurements can be
optimised to minimise the dominant theoretical uncertainties entering in the
determination of \Vub and \Vcb.

\section{Semileptonic Decays with Charm}
\label{sec:vcb}

The rate for inclusive \Bxclnu decays can be calculated in terms of 
the Heavy Quark Expansion (HQE), which is in powers of
$\alpha_s$ and $1/m_b$:
\begin{equation}
  \label{eq:vcbincl}
  \Gamma_{sl} = \frac{G_F m_b^5}{192 \pi^3} |V_{cb}|^2 (1+A_{EW})A_{pert}(\alpha_s)
  A_{non-pert}(1/m_b, 1/m_c, a_i),
\end{equation}
where $A_{EW}$ and $A_{pert}$ represent electroweak and QCD
perturbative corrections, respectively. The non-perturbative QCD part,
$A_{non-pert}$, is expanded in terms of the heavy quark masses ($m_b$,
$m_c$), with $a_i$ as coefficients. 
Buchm{\"u}ller and Fl{\"a}cher\cite{ref:HQEfit} combined the
measurements of moments of lepton energy $(E_\ell)$ and hadronic final
state invariant mass $(m_X)$ spectra from various experiments. In
addition they also included moments of photon energy spectrum from
radiative \btosgam decays to increase the sensitivity to some
parameters such as $m_b$.
A global fit to heavy-quark parameters gives 
$\Vcb = (41.96 \pm 0.23 \pm 0.35 \pm 0.59) \cdot 10^{-3}$ and 
$m_b = (4.59\pm0.04)\gev$.

The measurement of exclusive \Bxclnu decays is a good cross check of
the inclusive measurements and an important measurement by
itself. In this case, the main theoretical uncertainty
comes from the assumptions about the form factors.
The technique of determining \Vcb by using \Btodstlnu decays is well
established. The differential distribution can be written in terms of
$w$, the $D^*$ boost in the \B rest frame, as
\begin{equation}
  \frac{d\Gamma(\Btodstlnu)}{dw} = \frac{G_F^2 \Vcb^2}{48 \pi^3} {\cal{F}}(w)^2 {\cal{G}}(w)  
\end{equation}
where ${\cal{G}}(w)$ is a phase space factor and ${\cal{F}}(w)$ is a
form factor (FF). ${\cal{F}}(1)=1$ in the heavy quark limit, but lattice
QCD\cite{ref:hashimoto} can be used to compute effects due to finite
quark masses, leading to ${\cal{F}}(1)=0.919^{+0.030}_{-0.035}$.  The
shape of ${\cal{F}}(w)$ cannot be predicted by theory, and is
parameterised in terms of a slope $\rho^2$ and FF ratios
$R_1$ and $R_2$, independent of $w$. The helicity amplitudes entering
in the \Btodstlnu decay are functions of the above parameters. These
amplitudes can be determined by fitting the fully differential
rate of \Btodstlnu decays as a function of $w$ and three angles.
\babar\cite{ref:mandeep} measured the FF parameters
following this approach, giving $R_1 = 1.396 \pm 0.074$, $R_2
= 0.885 \pm 0.048$ and $\rho^2 = 1.145 \pm 0.075$, where the errors
are a factor 5 better than in previous
determinations\cite{ref:CLEOFF}. 
Using the improved values of $R_1$ and $R_2$ the \babar
exclusive \Vcb measurement\cite{ref:BBRexclvcb} gives
$\Vcb = (37.6 \pm 0.3_{stat} \pm 1.3_{syst} \pm 1.4_{theory}) \cdot 10^{-3}$.

\section{Charmless Semileptonic Decays}
\label{sec:vub}

\Vub is related to the full rate of inclusive \Bxulnu
decays by an expression equivalent to
Eq.~\ref{eq:vcbincl}, giving a theory uncertainty of $\sim5\%$. 
In practise, the accessible rate
is much reduced and the theoretical uncertainty increases
considerably, since the overwhelming background from
\Bxclnu\ decays must be suppressed by stringent kinematic
requirements. These cuts are all based on the mass difference between
$u$ and $c$ quark. The distributions of
$E_\ell$ and $q^2$, the squared invariant mass of the lepton pair,
extends to higher values for signal, whereas the $m_X$ spectrum and 
the light-cone momentum $P_+= E_X - |\vec{p}_X|$ are
concentrated at lower values. Therefore regions of the phase space can
be selected where the signal over background ratio is
adequate. However, in some regions HPE breaks down and a so-called shape
function is needed to resum non-perturbative contributions.
It depends on $m_b$ and heavy quark parameters,
and most of the theoretical uncertainty in inclusive \Vub
determinations is due to our imperfect knowledge of them.
\Vub is determined from the measurement of the charmless semileptonic
partial branching fraction, $\Delta {{\cal{B}}(\Bxulnu)}$, the \B
meson lifetime, $\tau_b$, and the rate $\zeta(\Delta\Phi)$, which is
predicted by theory (BLNP\cite{ref:blnp}, DGE\cite{ref:gardi}) and
depends on the phase space region, $\Delta\Phi$, defined by kinematic
cuts:
\begin{equation}
  \Vub = \sqrt{\frac{\Delta {\cal{B}}(\Bxulnu)}{\tau_b \cdot \zeta(\Delta\Phi)}}.
\end{equation}

The \B-factories have studied several kinematic variables to measure
\Vub.
The endpoint of the lepton energy spectrum for charmless decays is
well above the one for charm decays ($E_\ell>2.3\gev$). The good
knowledge of the charm background allows to push this cut below the
charm threshold, thereby increasing the acceptance and decreasing
theory uncertainty. The
results\cite{ref:babar-endpoint,ref:belle-endpoint,ref:cleo-endpoint}
are summarised in Tab.~\ref{tab:inclvub}.
Reconstructing unambiguously other variables involving
either the neutrino or the \X system is experimentally challenging and
requires more knowledge of the whole event. This can be achieved by
reconstructing one \B in a pure hadronic mode and studying the
recoiling \B, whose momentum and flavour are then known. 
This technique provides signal
over background ratios of about one or higher, at the expense of a very
small signal efficiency (${\cal{O}}(10^{-3})$).
Belle has measured \Vub for three different combinations of
kinematical variables, shown in Tab.~\ref{tab:inclvub}. 
The \babar result for $m_x$-\Q agrees within errors with the Belle
measurement.
The latest average from HFAG\cite{ref:hfag} using
BLNP\cite{ref:blnp} gives $\Vub=(4.45 \pm 0.20 \pm 0.26)\cdot 10^{-3}$. Using the
alternative approach by DGE\cite{ref:gardi} gives $\Vub=(4.41 \pm 0.20 \pm
0.20)\cdot 10^{-3}$.  Both results agree very well and the total uncertainty on \Vub
is 7.4\%.


The differential rate for exclusive \Bxulnu decays in
terms of \Q is proportional to $\Vub^2 F(\Q)^2$. In the
simple case of \Btopilnu and massless leptons only one FF is needed.
The absolute value of this FF is predicted by several theoretical
frameworks (light-cone sum rules (LCSR)\cite{Ball:2004ye}, lattice
QCD (LQCD)\cite{Shigemitsu:2004ft,Okamoto:2004xg}, and quark models); the
dependence on \Q can be checked experimentally, thereby allowing to
discriminate different theoretical calculations.
The \babar measurement\cite{Aubert:2005cd} agrees with 
LCSR and LQCD but disfavours the quark model ISGWII\cite{Scora:1995ty}.
The total branching ratio is $(1.38 \pm 0.10 \pm 0.16
\pm 0.08) \cdot 10^{-4}$, where the errors are statistical, 
systematic, and due to FF shape uncertainties.
Using LQCD\cite{Okamoto:2004xg}, this translates into $\Vub = (3.82 \pm 0.14 \pm 0.22 \pm
0.11^{+0.88}_{-0.52})\cdot 10^{-3}$, where the fourth error reflects the
uncertainty of the FF normalisation.
Belle performed a similar measurement\cite{Abe:2005ie} using
\Btodstlnu tagged events giving consistent results.

\begin{table}[t]
  \begin{center}
    \begin{tabular}{|l|c|c|c|}
      \hline
      accepted region & $f_u$ 
      & $\Delta{\cal{B}} [10^{-4}]$ & $\Vub [10^{-3}]$\\
      \hline
      \babar ($E_e>2.0\,\gev$)\cite{ref:babar-endpoint}
      & 0.26     & $5.3\pm 0.3\pm 0.5$           & $4.41\pm 0.29\pm 0.31$ \\ 
      BELLE ($E_e>1.9\,\gev$)\cite{ref:belle-endpoint}
      & 0.34     & $8.5\pm 0.4\pm 1.5$           & $4.82\pm 0.45\pm 0.30$ \\ 
      CLEO ($E_e>2.1\,\gev$)\cite{ref:cleo-endpoint}
      & 0.19	   & $3.3\pm 0.2\pm 0.7$           & $4.09\pm 0.48\pm 0.36$ \\ 
      \babar ($m_X<1.7\,\gev, q^2>8\,\gev^2$)\cite{ref:babar-q2mx}
      & 0.34     & $8.7\pm 0.9\pm 0.9$           & $4.75 \pm 0.35 \pm 0.32$ \\ 
      BELLE ($m_X<1.7\,\gev, q^2>8\,\gev^2$)\cite{ref:belle-mx}
      & 0.34     & $8.4\pm 0.8\pm 1.0$           & $4.68 \pm 0.37 \pm 0.32$ \\
      BELLE ($P_+<0.66\,\gev$)\cite{ref:belle-mx}
      & 0.57     & $11.0\pm 1.0\pm 1.6$          & $4.14 \pm 0.35 \pm 0.29 $ \\
      BELLE ($m_X<1.7\,\gev$)\cite{ref:belle-mx}
      & 0.66     & $12.4\pm 1.1\pm 1.2$          & $4.06 \pm 0.27 \pm 0.24$ \\ 
      \hline
    \end{tabular}
    \caption{Measurements of partial branching
      fractions $\Delta\cal{B}$ for inclusive \Bxulnu\ decays and \Vub, 
      adjusted by HFAG to common input parameters. $f_u$ is the space phase acceptance. 
      The errors on \Vub refer to experimental and theoretical
      uncertainties, respectively. 
      \label{tab:inclvub}}
  \end{center}
\end{table}

\section{Summary}

Inclusive measurements of \Vcb give a precision of 2\% dominated by
HQE theory uncertainties. Inclusive measurements of \Vub have reached
a precision of 7\% dominated by HQE parameters.
The exclusive measurements provide important cross checks and give
consistent results within the still large FF uncertainties.

\appendix

\def\Journal#1#2#3#4{{#1} {\bf #2}, #3 (#4)}
\def\NCA{\em Nuovo Cimento}
\def\NIM{\em Nucl. Instrum. Methods}
\def\NIMA{{\em Nucl. Instrum. Methods} A}
\def\NPPS{{\em Nucl. Phys. Proc. Suppl.}}
\def\NPB{{\em Nucl. Phys.} B}
\def\PLB{{\em Phys. Lett.}  B}
\def\PRL{\em Phys. Rev. Lett.}
\def\PRD{{\em Phys. Rev.} D}
\def\PRDRC{{\em Phys. Rev.} D-RC}
\def\ZPC{{\em Z. Phys.} C}
\def\EPJ{{\em Eur. Phys. J.} C}
\def\JHP{\em JHEP}
\def\IJMP{{\em Int. J. Mod. Phys} A}



\begin{thebibliography}{99}

\bibitem{ref:HQEfit}
O.\ Buchm\"uller and H.\ Fl\"acher, \Journal{\PRD}{73}{073008}{2006}. 

\bibitem{ref:hashimoto} S. Hashimoto {\it et al.\ }, \Journal{\PRD}{66}{014503}{2002}. 
\bibitem{ref:mandeep} B.\ Aubert {\it et al.\ }(BABAR Collab.), hep-ex/0602023.

\bibitem{ref:CLN} I.\ Caprini, L.\ Lellouch, and M.\ Neubert, \Journal{\NPB}{530}{153}{1998}. 
\bibitem{ref:CLEOFF} J.E.\ Duboscq {\it et al.\ }(CLEO Collab.), 
\Journal{\PRL}{76}{3898}{1996}.
\bibitem{ref:BBRexclvcb} B.\ Aubert {\it et al.\ }(BABAR Collab.), 
\Journal{\PRDRC}{71}{051502}{2005}. 

\bibitem{ref:blnp}
B.O.\ Lange, M.\ Neubert, G.\ Paz, \Journal{\PRD}{72}{073006}{2005}. 
\bibitem{ref:gardi}
J.R.\ Andersen and E.\ Gardi, \Journal{\JHP}{0601}{097}{2006}.
%
\bibitem{ref:babar-endpoint} B.\ Aubert {\it et al.\ }(BABAR Collab.), 
  \Journal{\PRD}{73}{012006}{2006}. 
\bibitem{ref:belle-endpoint} A.\ Limosani {\it et al.\ }(BELLE Collab.)
  \Journal{\PLB}{621}{28}{2005}. 
\bibitem{ref:cleo-endpoint} A.\ Bornheim {\it et al.\ }(CLEO Collab.), 
  \Journal{\PRL}{88}{231803}{2002}.
\bibitem{ref:hfag} E.\ Barberio {\it et al.\ }(Heavy Flavor Averaging Group), 
  hep-ex/0603003. 
\bibitem{ref:babar-q2mx} B.\ Aubert {\it et al.\ }(BABAR Collab.), hep-ex/0507017.  
\bibitem{ref:belle-mx} I.\ Bizjak {\it et al.\ }(BELLE Collab.),
\Journal{\PRL}{95}{241801}{2005}. 

\bibitem{Ball:2004ye} P.~Ball and R.~Zwicky,
  \Journal{\PRD}{71}{014015}{2005}.
\bibitem{Shigemitsu:2004ft} J.~Shigemitsu {\it et al.},
  \Journal{\NPPS}{140}{464}{2005}.
\bibitem{Okamoto:2004xg} M.~Okamoto {\it et al.},
  \Journal{\NPPS}{140}{461}{2005}.
\bibitem{Aubert:2005cd} B.~Aubert {\it et al.} (BABAR Collab.),
  \Journal{\PRD}{72}{051102}{2005}. 
\bibitem{Scora:1995ty} D.~Scora and N.~Isgur, 
  \Journal{\PRD}{52}{2783}{1995}.
\bibitem{Abe:2005ie} K.~Abe {\it et al.} (BELLE Collab.), hep-ex/0508018.

\end{thebibliography}
\end{document}